\documentclass[prd,twocolumn,superscriptaddress,preprintnumbers,nofootinbib]{revtex4}

\pdfoutput=1

\usepackage{amsfonts,amssymb,amsmath}
\usepackage{bm}
\usepackage{bbm}
\usepackage{epsfig}

\newcommand{\la}{\lambda}

\newcommand{\cL}{\mathcal{L}}
\newcommand{\cO}{\mathcal{O}}

\newcommand{\Z}{\mathbb{Z}}
\newcommand{\U}{\mathrm{U}}
\newcommand{\SUC}{\mathrm{SU(3)_C}}
\newcommand{\SUL}{\mathrm{SU(2)_L}}
\newcommand{\UY}{\mathrm{U(1)_Y}}

\newcommand{\cc}{\mathrm{c.c.}}

\newcommand{\tev}{~\mathrm{TeV}}
\newcommand{\gev}{~\mathrm{GeV}}
\newcommand{\pb}{~\mathrm{pb}}

\newcommand{\sla}[1]{
   \setbox0=\hbox{$#1$} \dimen0=\wd0 
   \setbox1=\hbox{/} \dimen1=\wd1
   \ifdim\dimen0>\dimen1                            
      \rlap{\hbox to \dimen0{\hfil/\hfil}}          
      #1                                            
   \else                                            
      \rlap{\hbox to \dimen1{\hfil$#1$\hfil}}       
      /                                             
   \fi}          
                                   
\begin{document}

\title{A unified, flavor symmetric explanation for the $t\bar{t}$ asymmetry and $Wjj$ excess at CDF}

\author{Ann E.~Nelson}
\affiliation{Department of Physics, University of Washington, Seattle, WA 98195} 
\author{Takemichi Okui}
\affiliation{Department of Physics, Florida State University, Tallahassee, FL 32306} 
\author{Tuhin S.~Roy}
\affiliation{Department of Physics, University of Washington, Seattle, WA 98195} 

\begin{abstract}
We present a simple, perturbative, and renormalizable model with a flavor symmetry which can explain 
both the $t\bar{t}$ forward-backward asymmetry and the bump feature present in the dijet mass 
distribution of the $W+jj$ sample in the range $120$--$160\gev$ that was recently reported by the 
CDF collaboration.  The flavor symmetry not only ensures the flavor/CP safety of the model, but also 
relates the two anomalies unambiguously.  It predicts a comparable forward-backward  asymmetry in 
$c\bar{c}$.  The  forward-backward asymmetry in  $b\bar{b}$ is, however, small.  A bump in the dijet 
mass distribution in $Z+jj$ sample is also predicted but with a suppressed cross-section.
\end{abstract}

\maketitle

\section{Introduction}
Recently, the CDF collaboration has reported two interesting anomalies --- a large $t\bar{t}$ 
forward-backward (FB) asymmetry~\cite{CDFttbar2011, CDFttbar2011-dilep}, and a $3.2\sigma$ excess 
in the $120$--$160\gev$ range in the dijet mass distribution of the $W+jj$ sample~\cite{CDFbump} 
(see, however, Ref.~\cite{D0Wjj2011} for a D\O\ analysis). 
The recent report of the FB asymmetry also confirms the trend suggested by the earlier measurements 
by CDF~\cite{CDFttbar2008, CDFttbar2009} and D0~\cite{D0ttbar2007, Abazov:2011rq}.  

It is a straightforward exercise to fit these two anomalies by introducing new particles with 
appropriately chosen masses and couplings.  However, the nature of these anomalies suggests that 
the new physics should couple to standard-model (SM) quarks at tree level with an $\cO(0.1)$--$\cO(1)$ 
coupling along with a nontrivial quark flavor structure.  Such a new physics typically faces strong 
constraints from precision flavor and CP constraints, unless the model is equipped with a flavor
symmetry.
  
In this paper, we present a weakly-coupled, renormalizable field theory with a flavor symmetry to 
explain both anomalies. (For attempts to generate just the FB asymmetry preserving the full flavor 
symmetries, see e.g.~Refs.~\cite{Delaunay:2010dw,Delaunay:2011vv,Grinstein:2011yv, MartinZoltan}.)  We introduce just one multiplet 
of scalars with a single coupling to SM quarks dictated by the flavor symmetry. The flavor symmetry 
ensures that the only source of flavor and CP violations is $V_\text{CKM}$.  It also relates the 
sizes of the two anomalies in a definite manner, and entails additional predictions.  

The rest of the paper is organized as follows.  In Sec.~\ref{sec:model}, we define our model with 
an emphasis on the flavor symmetry structure, which keeps flavor/CP violations under control without 
tuning or an ad hoc choice of couplings.  In Sec.~\ref{sec:constraints}, we go through various 
potential constraints on the model for the values of parameters necessary for obtaining the 
$t\bar{t}$ FB asymmetry and $Wjj$ bump. Sec.~\ref{sec:FB} shows our estimation of the asymmetry,
while Sec.~\ref{sec:bump} shows the details of the bump feature in the dijet mass spectrum as 
predicted in our model.  Our concluding reflections and a brief discussion of various implications 
of the model are included in  Sec.~\ref{sec:conc}.

\section{The model}
\label{sec:model}
The flavor symmetry we propose is a subgroup of the $\U(3)^3$ quark flavor symmetry of the SM:
\begin{equation}
  \left( \prod_{i=1}^3 \U(1)_{q_{\mathrm{L}i}} \times \U(1)_{u_{\mathrm{R}i}} \right)\! 
  \times \U(3)_d \times \Z_3  \,,
\label{eq:symmetry}
\end{equation}
where $q_{\mathrm{L}i}$ and $u_{\mathrm{R}i}$ have charge $+1$ under $\U(1)_{q_{\mathrm{L}i}}$ and
$\U(1)_{u_{\mathrm{R}i}}$, respectively, 
while $d_\mathrm{R}$ is a $\mathbf{3}$ of $\U(3)_d$.  $\Z_3$ cyclically permutes the flavor indices 
of $q_{\mathrm{L}i}$ and $u_{\mathrm{R}i}$ ($i=1,2,3$), but not of $d_{\mathrm{R}i}$.  The lepton 
sector of our model is identical to that of the SM, and will not be discussed in this paper.

In the SM, one can always go to a basis where $Y_u$ is diagonal.  In the limit of neglecting both 
$Y_u$ and $Y_d$, the SM possesses the flavor symmetry~\eqref{eq:symmetry}.  Turning on the diagonal 
(but non-degenerate) $Y_u$ breaks the symmetry~\eqref{eq:symmetry} to its subgroup 
$\U(1)_{B_1} \times \U(1)_{B_2} \times \U(1)_{B_3} \times \U(3)_d$, where 
$q_{\mathrm{L}i}$ and $u_{\mathrm{R}i}$ have charge $+1$ under $\U(1)_{B_i}$.  This subgroup still forbids all flavor violations. Turning on $Y_d$ then breaks 
$\U(1)_{B_1} \times \U(1)_{B_2} \times \U(1)_{B_3} \times \U(3)_d$ down to the baryon number 
$\U(1)_B$, thus introducing flavor mixing.  However, since $Y_d$ breaks (and only $Y_d$ breaks) 
$\U(3)_d$,  we can always bring $Y_d$ into the form 
$Y_d = V_\text{CKM}\, \mathrm{diag}(y_d, y_s, y_b)$, ensuring that $V_\text{CKM}$ is the only source 
of flavor violation.  Also,  note that $\mathrm{diag}(y_u, y_c, y_t)$ and 
$\mathrm{diag}(y_d, y_s, y_b)$ can both be taken  to be positive definite, rendering $V_\text{CKM}$ 
the only source of CP violation.%
\footnote{We neglect the QCD vacuum angle.  It is straightforward to add an axion to our model to 
solve the strong CP problem.}

Our fundamental assumption is that this symmetry breaking pattern persists for new physics beyond 
the SM as well.  In other words, new physics should fully respect the flavor 
symmetry~\eqref{eq:symmetry} in the limit $Y_u, Y_d \to 0$, and  so $Y_u$ and $Y_d$ remain the only 
spurions breaking the symmetry~\eqref{eq:symmetry}.  This can be thought of as a variant of minimal 
flavor violation (MFV)~\cite{MFV1, MFV2, MFV3, MFV4, MFV5}, and, in particular, the breaking pattern
$\U(1)_{B_1} \times \U(1)_{B_2} \times \U(1)_{B_3} \times \U(3)_d \longrightarrow U(1)_B$ by $Y_d$
has been studied in the context of the supersymmetric SM~\cite{Barbieri1, Barbieri2}.%
\footnote{A crucial difference between MFV and our flavor symmetry breaking pattern 
is in the spurion structure in the up-quark sector.  
In MFV there are 9 complex spurions for the up-quark sector, that is, the 3$\times$3 matrix 
$Y_u$ transforming as $(\mathbf{3},\mathbf{\bar{3}})$ under 
$\mathrm{SU}(3)_{q_\mathrm{L}}\times\mathrm{SU}(3)_{u_\mathrm{R}}$, while
in our case there are only three real spurions, $(Y_u)_{ii}$ ($i=1,2,3$) carrying the charge
$(1,-1)$ for $\U(1)_{q_{\mathrm{L}i}} \times \U(1)_{u_{\mathrm{R}i}}$. By assumption $Y_d$
 is the only spurion which breaks CP or the $\U(1)_{B_{1,2,3}}$ quantum numbers.
Note  that the up-type quarks are in the mass basis at the outset  
and that, unlike in MFV,  unitary rotations done on  $q_\mathrm{L}$ or $u_\mathrm{R}$  are not  approximate symmetries of the Lagrangian.}  
We introduce a $\Z_3$ triplet of complex scalar fields, $\Phi=(\Phi_1, \Phi_2, \Phi_3)$, 
where the gauge quantum number of $\Phi$ is $(\mathbf{1}, \mathbf{2})_{-1/2}$ under 
$(\SUC, \SUL)_{\UY}$ representation. $\Phi_i$ ($i=1,2,3$) are singlets under $\U(3)_d$, but they 
are charged under 
$\U(1)_{q_{\mathrm{L}1}} \times \U(1)_{q_{\mathrm{L}2}} \times \U(1)_{q_{\mathrm{L}3}}$ as
\begin{equation}
  \Phi_1 \sim (0,0,1) \,,\>
  \Phi_2 \sim (1,0,0) \,,\>
  \Phi_3 \sim (0,1,0) \,.
\end{equation}
while under $\U(1)_{u_{\mathrm{R}1}} \times \U(1)_{u_{\mathrm{R}2}} \times \U(1)_{u_{\mathrm{R}3}}$ as
\begin{equation}
  \Phi_1 \sim (0,-1,0) \,,\>
  \Phi_2 \sim (0,0,-1) \,,\>
  \Phi_3 \sim (-1,0,0) \,,
\end{equation}
Note that these charge assignments respect $\Z_3$.  The tree-level Lagrangian reads:
\begin{equation}
\begin{split}
  & \cL_\text{tree} 
   = \cL_\text{SM} + (D_\mu\Phi)^\dag (D^\mu\Phi) - m_\Phi^2 \Phi^\dag\Phi  \\
   & \quad\>\>\> 
     -\la\, (\overline{q}_{\mathrm{L}1} \Phi_2 u_{\mathrm{R}3} 
             +\overline{q}_{\mathrm{L}2} \Phi_3 u_{\mathrm{R}1}
             +\overline{q}_{\mathrm{L}3} \Phi_1 u_{\mathrm{R}2} +\cc
            )  \\
   & \quad\>\>\> 
     -\xi\, (H^\dag \sigma^a H) (\Phi^\dag \sigma^a \Phi) -\xi' (H^\dag H) (\Phi^\dag \Phi)  \\
   & \quad\>\>\> 
     -\frac{\zeta}{4} (\Phi^\dag \Phi)^2  \,.
\label{eq:lagrangian}
\end{split}
\end{equation}
The new physics is completely invariant under the flavor symmetry~\eqref{eq:symmetry}. Therefore, 
the flavor symmetry~\eqref{eq:symmetry} is broken only by $Y_u$ and $Y_d$, just as in the SM\@.
Also, we have chosen $\la$ to be real without loss of generality, by redefining the phase of $\Phi$. 
No new CP phase has, therefore, been introduced in this model.

The tree-level Lagrangian~\eqref{eq:lagrangian} has three phenomenologically relevant parameters: 
$m_\Phi$, $\la$,  and $\xi$. $\Z_3$ dictates that all three components of $\Phi$ have the 
equal mass $m_\Phi$, and that they all couple to the SM quarks with the same strength $\la$ and 
likewise to the Higgs via $\xi$ and $\xi'$. The interesting role of $\xi$ is that it splits the 
masses of the neutral ($\Phi^0$) and charged ($\Phi^-$) components of $\Phi$ as:
\begin{equation}
  m_{\Phi^0}^2 = m_{\Phi,\text{eff}}^2 - \xi v^2  \,,\quad
  m_{\Phi^-}^2 = m_{\Phi,\text{eff}}^2 + \xi v^2  \,,
\end{equation}
where $v=174\gev$ and $m_{\Phi,\text{eff}}^2 \equiv m_{\Phi}^2 + \xi' v^2$.  We choose 
$m_{\Phi^0} = 160\gev$ and $m_{\Phi^-} = 220\gev$, which corresponds to 
$m_{\Phi,\text{eff}} = 192\gev$ and $\xi = 0.38$.  We take $\la$ to be $1.4$.  This might appear too 
large to keep $\la$ perturbative up to very high scale.  Fortunately, the one-loop RG equation for 
$\la$ is similar to that of the top Yukawa coupling in the SM and there is a quasi-fixed point 
near $\la \approx 1.4$. 

At loop level, counter-terms $\delta\cL$ must be added to the Lagrangian \eqref{eq:lagrangian} for 
renormalization.  We assume that all terms required to renormalize the theory are present in 
$\delta\cL$, at the minimal level required to avoid fine tuning.%
\footnote{We have the usual ``hierarchy problem'' for $m_H$ and $m_\Phi$ as well as other 
dimension-2 operators in $\delta\cL$.  In this paper, we simply tune $m_H$ and $m_\Phi$ and use 
dimensional regularization with (modified) minimal subtraction to obtain the natural sizes of 
dimension-2 operators in $\delta\cL$, but it is straightforward to supersymmetrize the model to 
justify this.}
This assumption is technically natural, and may be justified by assuming that our Lagrangian arises 
from a more fundamental theory in which $Y_u$ and $Y_d$ are the only parameters breaking the flavor 
symmetry~\eqref{eq:symmetry}.  As an example of terms in $\delta\cL$,  renormalization requires the 
counter-terms 
$\overline{q}_{\mathrm{L}1} Y_{u1} \Phi^\dag_3 (Y_d d_{\mathrm{R}})_2 
+ (\text{cyclic permutations})$ 
with a common coefficient $\sim \la/(16\pi^2)\, \log\Lambda$. So we include these operators in 
$\delta\cL$ 
with a single coefficient $\sim \la/(16\pi^2)$.  This also exemplifies the general principle that all 
generated operators respect the flavor symmetry~\eqref{eq:symmetry}, broken only by $Y_u$ and $Y_d$. 
Moreover, since no operators are generated with an independent phase, renormalization does not require 
introducing new phases.  The property that $V_\text{CKM}$ is the only source of flavor and CP 
violations, therefore, remains intact at the quantum level.

\section{Constraints}
\label{sec:constraints}

The property that $V_\text{CKM}$ is the only source of flavor/CP violations, and the fact that the 
mass scale for $\Phi^0$ and $\Phi^-$ is similar to or larger than  the top quark mass, imply that flavor/CP violations 
involving $\Phi$ is at most comparable to those in the SM\@.
For example, consider bounds from $D^0$-$\overline{D}{}^0$ mixing, that is, 
4-fermion operators with two $c$ and two $\bar{u}$ fields that arise upon integrating 
$\Phi$ out.  Recall that, in our flavor symmetry breaking pattern, the only flavor non-diagonal 
spurion is $Y_d = V_\text{CKM}\, \mathrm{diag}(y_d, y_s, y_b)$.
The flavor symmetry~(\ref{eq:symmetry}) thus dictates that the simplest combination of 
spurions that can change $c$ to $u$ must involve the combination 
$(Y_d^{\phantom\dag} Y_d^\dag)_{12}^{\phantom\dag} 
= (V_\text{CKM}\,\mathrm{diag}(y_d^2,y_s^2,y_b^2) V_\text{CKM}^\dag)_{12}^{\phantom\dag} 
\sim \cO(10^{-6})$.  Since this combination converts 
$c_\mathrm{L}$ to $s_\mathrm{L}$, the coefficient of $(\bar{u}_\mathrm{L} c_\mathrm{L})^2$ is 
on the order of $\bigl[(Y_d^{\phantom\dag} Y_d^\dag)_{12}^{\phantom\dag} / m_\Phi\bigr]^2 
\sim (10^5~\mathrm{TeV})^{-2}$ 
multiplied by certain powers of the couplings $\lambda$ and $g$ and the loop factor $1/(16\pi^2)$.
This is safely much smaller than the experimental 
bound~$\sim (10^3\>\mathrm{TeV})^{-2}$~\cite{Isidori:2010kg}.
For the $(\bar{u}_\mathrm{L} c_\mathrm{R}) (\bar{u}_\mathrm{R} c_\mathrm{L})$ operator, 
the spurions $(Y_u)_{22} \sim 10^{-2}$ and $(Y_u)_{11} \sim 10^{-5}$ have
to be further inserted to convert $c_\mathrm{L}$ to $c_\mathrm{R}$ and 
$u_\mathrm{L}$ to $u_\mathrm{R}$, respectively, rendering the coefficient
of the operator way below the bound~$\sim (10^4\>\mathrm{TeV})^{-2}$~\cite{Isidori:2010kg}.
Therefore, $D^0$-$\overline{D}{}^0$ mixing is not an issue at all in our model, 
thanks to the flavor symmetry.

The most stringent flavor bounds on our model arise from the 
4-fermion operators that are generated via the tree-level exchange of $\Phi$. In the gauge basis, 
these are
\begin{equation}
  (\bar{q}_{\mathrm{L}2} u_{\mathrm{R}}) (\bar{u}_{\mathrm{R}} q_{\mathrm{L}2}) \,,\>
  (\bar{q}_{\mathrm{L}3} c_{\mathrm{R}}) (\bar{c}_{\mathrm{R}} q_{\mathrm{L}3}) \,,\>
  (\bar{q}_{\mathrm{L}1} t_{\mathrm{R}}) (\bar{t}_{\mathrm{R}} q_{\mathrm{L}1}) \,.
\label{eq:4fermi}
\end{equation}
The first and second operators can contribute to hadronic $b$ decays.  In the mass basis, they contain 
\begin{equation}
  V_{cb}^* V_{ci} (\bar{b}_{\mathrm{L}} u_{\mathrm{R}}) (\bar{u}_{\mathrm{R}} d_{\mathrm{L}i}) + 
  V_{tb}^* V_{ti} (\bar{b}_{\mathrm{L}} c_{\mathrm{R}}) (\bar{c}_{\mathrm{R}} d_{\mathrm{L}i})  \,.
\label{eq:suckers}
\end{equation}
This comes from the tree-level exchange of $\Phi^-$, so its coefficient is $\la^2/m_{\Phi^-}^2$.
The first operator contributes to the charmless process $b\rightarrow s \bar{u} u$. The particle 
data book~\cite{Nakamura:2010zzi} specifies the total inclusive branching fraction of $B$ mesons into 
charmed modes to be $(95\pm5 )\%$. For $m_{\Phi^-}=220\gev$ and $\la = 1.4$, the leading order 
spectator decay model gives a branching fraction for the $b\rightarrow s \bar{u} u$ mode of $15\%$, 
which is within the $2\sigma$ margin of error. CP constraints do not pose a problem for the new 
contribution to decays. The CP phase in the new contribution to $b\rightarrow c \bar{c} s$ is almost 
the same as the phase of the standard model contribution. The phase of the new contribution to the 
$b\rightarrow s \bar{u} u$ mode is the same as the gluonic penguin contribution, which so far, 
is consistent with experiments. Of greater concern is the nonstandard increase in the hadronic width. 
However the heavy quark expansion, which is needed for a theoretical computation of the hadronic 
width~\cite{Altarelli:1991dx,Bagan:1994qw,Czarnecki:2005vr}, may have significant uncertainty for 
this computation~\cite{Dighe:2007gt,Bauer:2010dga}.  Should further experimental tests of $B$ decay 
modes exclude such a large new contribution to hadronic $B$ decays then the $\Phi^-$ mass would have 
to be increased. 

The correction to the electroweak parameter $\alpha T$ is given by
\begin{equation}
\hspace{-0.1ex}  
  \alpha T\!   
  = \frac{3}{32\pi^2 v^2} 
     \!\!\left[ m_{\Phi^0}^2 + m_{\Phi^-}^2 
             -\frac{2 m_{\Phi^-}^2 m_{\Phi^0}^2}{m_{\Phi^-}^2 - m_{\Phi^0}^2} 
              \log\frac{m_{\Phi^-}^2}{m_{\Phi^0}^2}
         \right]\!
\label{eq:Tparam}
\end{equation}
where $\alpha$ is to be evaluated at the weak scale.
For $m_{\Phi^0} = 160\gev$ and $m_{\Phi^-}=220\gev$, we get $\alpha T = 1.5 \times 10^{-3}$. From the 
particle data book~\cite{Nakamura:2010zzi}, for a Higgs mass of 117(300) GeV, the $T$ parameter is 
constrained to be $T=0.07(0.16)\pm0.08$. Thus, our model is consistent with precision electroweak 
constraints without any tuning. However, the $T$ parameter contribution gives an upper bound on the 
mass of the $\Phi^-$.

$\Phi$ couplings to leptons are highly loop-suppressed; so precision lepton measurements (e.g.~the 
muon $g-2$) do not place constraints on our model. Since all components of $\Phi$ are unstable even 
in the collider time scale,  there are no cosmological constraints. Hence, we concentrate on the 
collider physics constraints below.%
\begin{itemize}
\item{The total $t\bar{t}$ production cross-section is not changed significantly.  Also note that the 
precise value of the theoretical prediction is still open to debate. The NLO+NNL calculations quote 
$\sim 10\%$ uncertainty~\cite{Moch:2008qy,Cacciari:2008zb,Kidonakis:2008mu, Moch:2008ai} and the 
resummations of threshold logs result in a smaller value~\cite{Ahrens:2010zv, Ahrens:2011mw}. When 
compared to the SM leading order result, we find that $d \sigma_{t \bar t}/d M_{t\bar{t}}$, where 
$M_{t\bar{t}}$ is the   invariant mass of the $t\bar{t}$ pair, in our model slightly decreases near 
the threshold but slightly increases at higher values of $M_{t\bar{t}}$. (See Sec.~\ref{sec:FB} for 
details).
}
\item{Single top production via $\Phi$ is suppressed.  The leading single top production comes from 
$b \bar{c} \to \Phi^-_1 \to W^- \Phi^0_1 \to W^- t \bar{c}$, which is highly  suppressed by the 
smallness of the $b$ and $c$ parton distribution functions (PDFs).  It is also suppressed by the 
3-body phase space if the $W^-$ is on-shell and the $\Phi^-$ is off-shell.  If the $\Phi^-$ is 
on-shell, then the $W^-$ must be off-shell, becoming a 4-body process. Either case, it is clearly 
much smaller than the SM counterpart $u\bar{d} \to W^+ \to t\bar{b}$, which is only suppressed by 
the offshellness of the $W^+$.  
}
\item{The CDF collaboration has searched for events with same-sign lepton pairs and at least one $b$ 
jet,  and found $3$ such events in $2~\mathrm{fb}^{-1}$ of data \cite{ditop-CDF}, where they expect 
$\sim 2$ events from background.  Di-top ($tt$) production can give such final states and is, thus, 
severely constrained. In our model, however, di-top production is extremely suppressed since
in the limit of neglecting $Y_d$, baryon numbers for the three up-type quarks are separately 
conserved as in Eq.~\eqref{eq:symmetry}.  Therefore, this is not a constraint for us.
}
\item{The top-quark width is not modified significantly.  For $m_{\Phi^0} = 160\gev$ and $\la=1.4$, 
the total top width is $1.6\gev$, which is well within the experimental 
limit~\cite{Abazov:2010tm,Aaltonen:2010ea}.
}
\item{There is a sizeable dijet production in our model via $\Phi_3$ exchange in the 
$s$- or $t$-channel.  For masses as low as $160\gev$ or $220\gev$, Tevatron has large SM dijet 
backgrounds due to the gluon
PDF's increasing rapidly 
at low parton $x$.  Thus, there are no constraints from Tevatron \cite{dijets-CDF}. 
The strongest bound comes from the CERN SPS collider ($p\bar{p}$ at $\sqrt{s} = 630\gev$).  The UA2 
collaboration at the SPS has placed $90\%$ C.L.~bounds $\approx 100\pb$ on a dijet resonance at 
$160\gev$, and $\approx 10\pb$ at $220\gev$~\cite{UA2}.  Among our dijet channels, those 
from $t$-channel $\Phi_3$ exchange give a smooth dijet mass distribution on top of the smooth 
huge SM background. So they could not have been picked up by the UA2 search.  For the $s$-channel 
contributions from $\Phi_3^0$ and $\Phi_3^-$, we find the cross-sections to be $\simeq 56\pb$ 
for $\Phi_3^0$ and $\simeq 15\pb$ for $\Phi_3^-$, with $\la = 1.4$.  The latter may appear to 
be in conflict with the UA2 $90\%$ C.L.~bound.  However, note that the UA2 bounds are 
for a {\em narrow} 
resonance such as $W'$ which can distinguish itself from the smooth SM dijet background. Our 
$\Phi_3^-$ resonance, on the other hand, is not narrow --- its width is quite large, $\approx 26\gev$ 
for $m_{\Phi^-}  = 220\gev$ and $\la=1.4$, which is expected to be smoothed out even further once 
parton showering and detector effects are taken into account.  Thus, the search optimized for a very 
narrow resonance has a reduced sensitivity to our $\Phi_3^-$.  
Furthermore, note that UA2 performed their analysis in 
the early days of QCD jet studies.  Their answer depends crucially on the quality of the Monte Carlo 
and the detector simulation which are primitive by today's standard.  They also use events with two 
exclusive jets, where jets were constructed using an infrared unsafe jet 
algorithm~\cite{Kunszt:1992tn}. We believe, therefore, that there are considerable uncertainties 
associated with their bounds, and that it is fair to regard the UA2 $90\%$ C.L.~bound as an 
order-of-magnitude limit. 
}
\end{itemize}

\section{The $t\bar{t}$ forward-backward asymmetry}
\label{sec:FB}
Neglecting the CKM mixings and non-valence partons in the incoming $p$ and $\bar{p}$, the relevant 
interactions for this are:
\begin{equation}
  \cL_\text{int}
  = -\la ( \bar{u}_\mathrm{L} \Phi^0_2 t_\mathrm{R}
          + \bar{d}_\mathrm{L} \Phi^-_2 t_\mathrm{R} +\cc)  \,.
\end{equation}
The $t\bar{t}$ forward-backward asymmetry arises at the Tevatron from the processes 
$u\bar{u} \to t\bar{t}$ with a $t$-channel $\Phi^0_2$ exchange, and $d\bar{d} \to t\bar{t}$ 
with a $t$-channel $\Phi^-_2$ exchange. 

A dedicated simulation including parton showering and detector effects is beyond the scope of this 
paper, partly due to the uncertainties in the SM prediction and the experimental measurement of the 
asymmetry. We simply perform our analysis at the parton level, and show that the asymmetry is 
generated with the right sign and is of the same order in magnitude. 

\begin{figure}[t]
\includegraphics[height=2in]{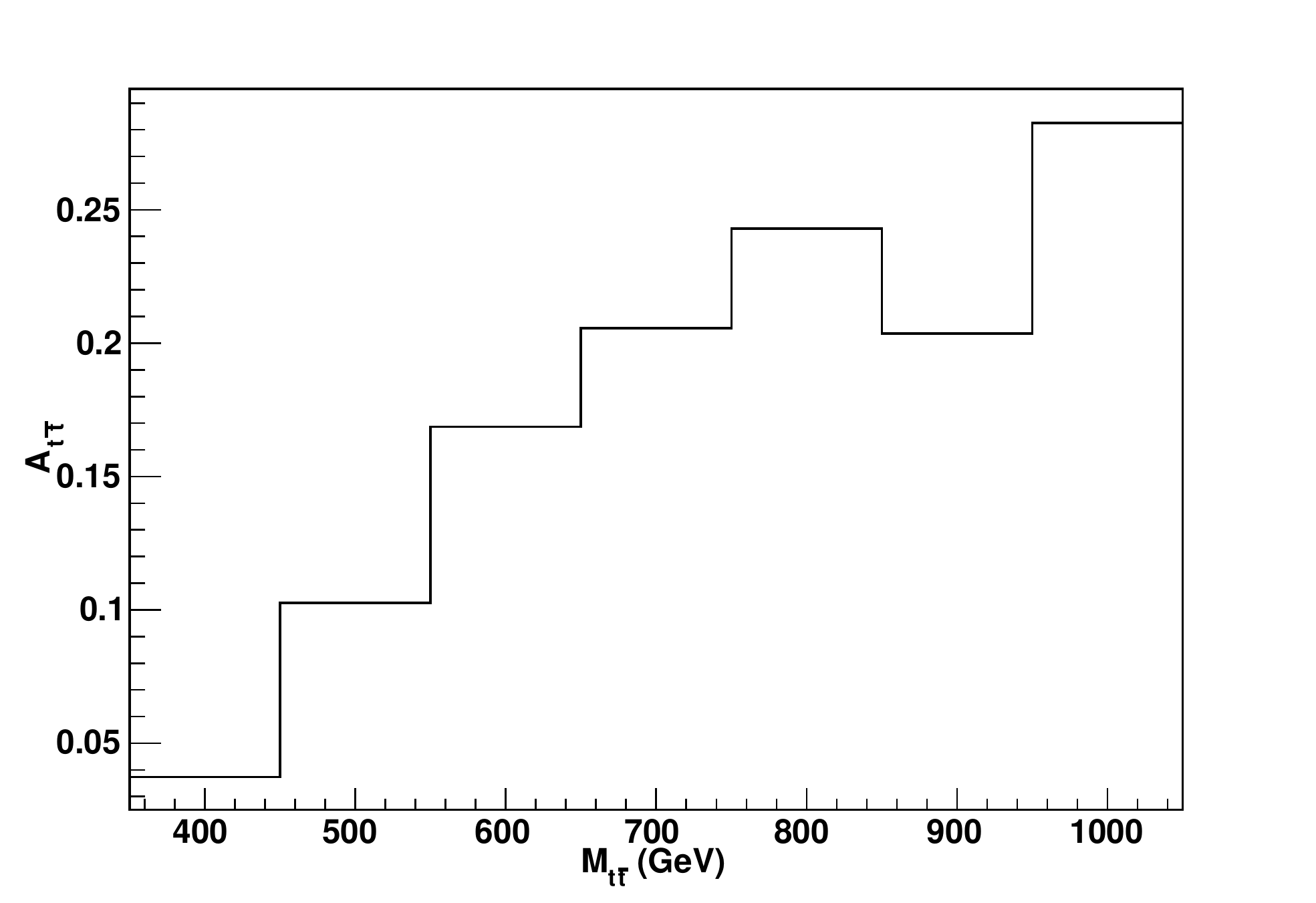}
\caption{$A_{t\bar{t}}$ as a function of the invariant mass of the $t\bar t$ pair, $M_{t\bar t}$. 
We calculated $ A_{t\bar{t}}$ after demanding both the tops to be within $\eta = \pm 2.0$.}
\label{fig:afb}
\end{figure}

We define the asymmetry as
\begin{equation}
  A_{t\bar{t}} =  \frac{N(\Delta y > 0) - N(\Delta y < 0)}{N(\Delta y > 0) + N(\Delta y < 0)} \, ,
\end{equation}
where $\Delta y$ is the rapidity difference between the $t$ and $\bar{t}$. Our $t\bar{t}$ sample was 
generated using {\tt Madgraph v4.4.48}~\cite{Alwall:2007st}.  We have imposed the following cuts on 
the $t\bar{t}$ pairs: $|\eta_t|, |\eta_{\bar{t}}| < 2.0$, and  $M_{t\bar{t}} > 450\gev$.  We find the 
asymmetry to be  $A_{t \bar t} \simeq 0.13$. Note that the asymmetry in our model does not depend 
linearly on $\cos\theta$ as it is assumed in Ref.~\cite{CDFttbar2011} to extrapolate the asymmetry 
to the full $4\pi$ solid angle. This is because our asymmetry is generated by $t$-channel exchange of 
particles similar in mass to the top quark, not from an $s$-channel heavy resonance.  It should, 
thus, be compared to the value $A_{t\bar{t}} \simeq 0.212\pm0.096$ for the reconstructed 
$M_{t\bar{t}} > 450\gev$ as actually measured within the CDF detector 
coverage~\cite{CDFttbar2011-dilep}.  

\begin{figure}[t]
\includegraphics[height=2in]{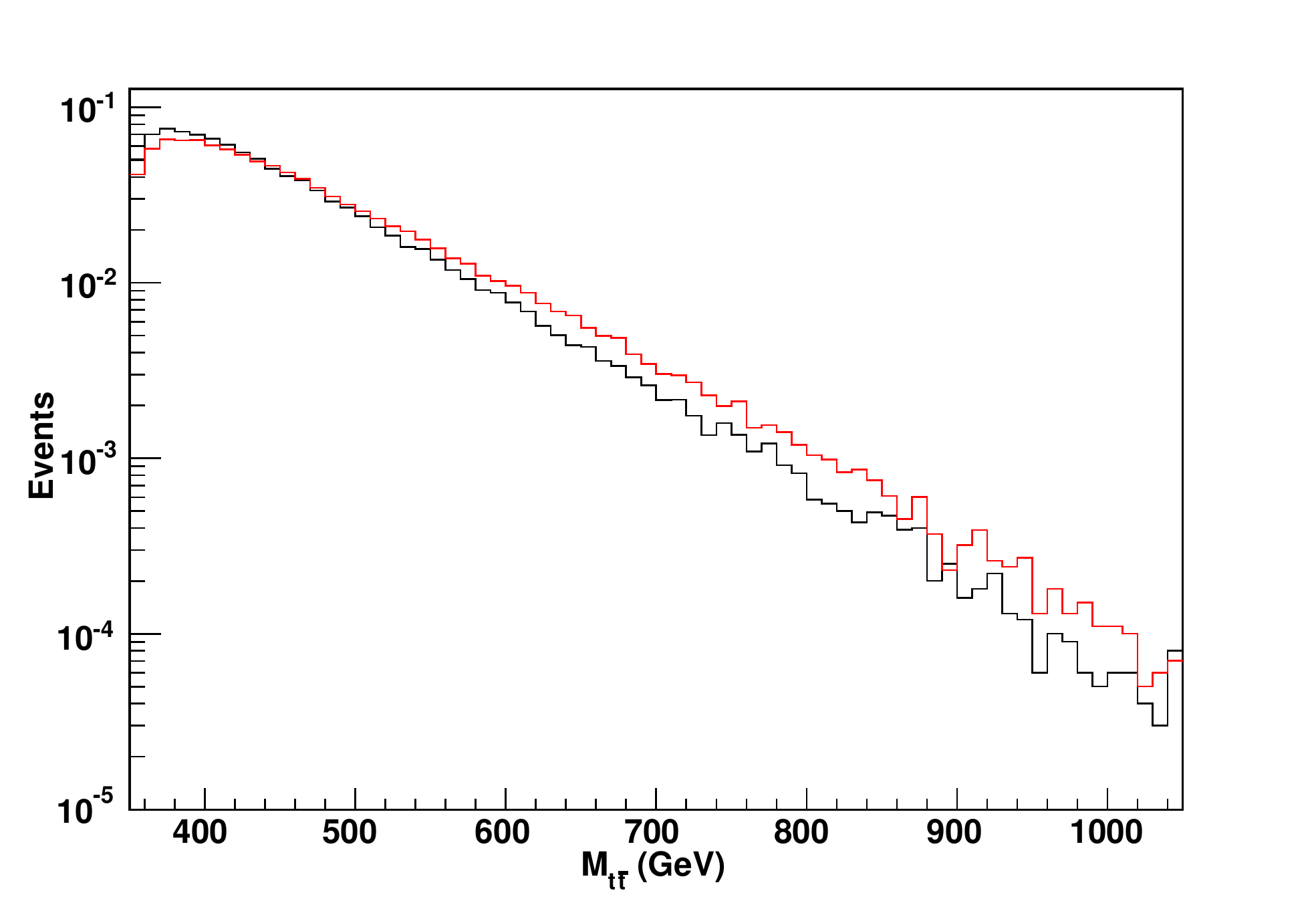}
\caption{Comparison of the area normalized distribution of $\sigma_{t\bar{t}} $ in the SM (in black) 
and in our model (in red). In both cases the cross-section is calculated at the leading order using 
{\tt Madgraph v4.4.48}.}
\label{fig:sigma_tt}
\end{figure}

We also check the total cross section of $t \bar{t}$ production at tree level. It is found to be 
within $10\%$ of the value as calculated in the SM at the leading order. Assuming the same $k$-factor 
as in the case of  the SM, we predict the total $t \bar t$ cross-section within theoretical 
uncertainties~\cite{Moch:2008qy,Cacciari:2008zb,Kidonakis:2008mu, Moch:2008ai}.  Deviation is seen when we check the cross-section as a function $M_{t \bar t}$.  In Fig.~\ref{fig:sigma_tt} we have 
compared the tree level cross-section, as calculated in our model, to that in the SM\@.  As shown in 
Fig.~\ref{fig:sigma_tt} and as reported in Refs.~\cite{AguilarSaavedra:2011vw}, the deviation is too small to give clear signal especially in early LHC data.  Note that we 
cannot make direct comparison to the experimental data, since higher
order corrections may change the shape of the curve besides the
overall cross-section.     
%

\section{The CDF excess in $Wjj$}
\label{sec:bump}
The relevant interactions for these processes are:
\begin{equation}
  \cL_\text{int}
  = -\la (\bar{c}_\mathrm{L} \Phi^0_3 u_\mathrm{R} 
          +\bar{s}_\mathrm{L} \Phi^-_3 u_\mathrm{R} +\cc)  \,.
\end{equation}
$Wjj$ final states via an intermediate $\Phi$ then dominantly arise at the Tevatron from the 
processes $u\bar{s} \to W^+ \Phi^0_3 \to W^+ u\bar{c}$ and its charge-conjugated process.

For $\la = 1.4$, the $Wjj$ signal cross-section is found to be $2.1\pb$. We use  
{\tt Madgraph v4.4.48} to generate the signal events, which are  subsequently   decayed, showered, 
and finally hadronized by {\tt PYTHIA v6.4}~\cite{Sjostrand:2006za}. We group the hadronic output 
of {\tt PYTHIA} into ``cells'' of size $\Delta \eta \times \Delta \phi = 0.1\times 0.1$\@. We sum 
the four momentum of all particles in each cell and rescale the resulting three-momentum such as to 
make cells massless.  Jet clustering is done using the CDF version of Run-II iterative cone algorithm 
with midpoint seeds, as implemented in {\tt Fastjet}~\cite{Cacciari:2005hq}. We do not perform any 
realistic detector simulation. 

\begin{figure}[t]
\includegraphics[height=2in]{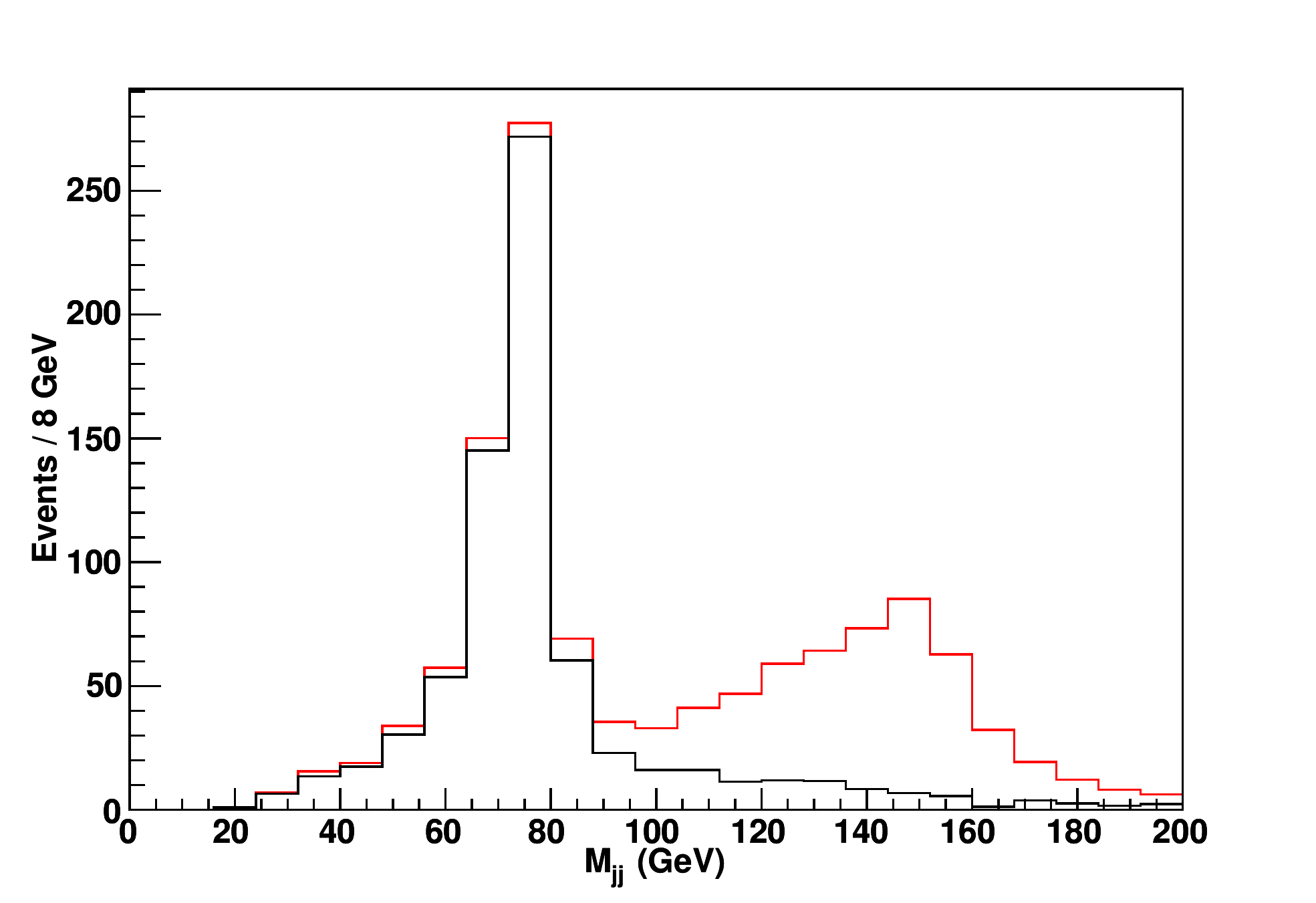}
\caption{The dijet invariant mass distribution. The red curve is due to the resonances in our model 
at $160\gev$. The black curve shows the same for SM diboson events. Note that the center of the peak 
is shifted to $150\gev$.}
\label{fig:Mjj}
\end{figure}

We demand exactly one isolated lepton with $p_\mathrm{T} > 20\gev$ and $|\eta| < 1.0$ plus missing 
transverse energy $\sla{E}_\mathrm{T} > 25\gev$.  The event also contains exactly two jets, each with 
$E_\mathrm{T} > 30\gev$ and within $|\Delta \eta| < 2.5$ of each other.  The dijet invariant mass 
distribution is plotted in Fig.~\ref{fig:Mjj}.  In order to get an idea of the size of the bump in 
our model, we have also plotted SM diboson events ($ZW^{\pm}, W^{+}W^{-}$) that pass all our selection 
criteria.  Note that the bump has all the features as seen in the data, including the size and the 
position of the peak at $145$--$150\gev$. 

The CDF has used a Gaussian profile function of narrow width to fit the excess. On the contrary, a 
quick glance at the the dijet mass distribution in data suggests that data favors a broader peak. The 
dijet mass distribution in the range $170$--$220\gev$ is characterized by upward fluctuations in 
each bin. Under modest smearing, we find that the position and the shape of the peak in our signal 
events remain relatively unaltered and slight excesses are generated in higher bins.  A true 
comparison, however, can only be made after a decent detector simulation, which is beyond the scope 
of this paper.  

One might expect that we should also see a similar excess in $Zjj$ sample at or around 
$120$--$160\gev$.  Interestingly, $Zjj$ production via an intermediate $\Phi^0$ dominantly arises 
from $u\bar{c} \to Z \Phi^0_3 \to Z u\bar{c}$. This process is, however, suppressed because of the 
$c$ PDF, leading to a much smaller cross-section $\simeq 0.3$~pb, consistent with the absence of 
observation of such an excess around $160\gev$.  For the $u\bar{s}$ initial state, the process must 
be $u\bar{s} \to Z \Phi^+_3 \to Z u\bar{s}$, which would lead to an excess in a different region 
(around $220\gev$). However, the cross-section for it is smaller ($\simeq 0.24$~pb) and 
the peak would appear even broader.

\section{Conclusions}
\label{sec:conc}
The $t\bar{t}$ forward-backward asymmetry and the $3.2\sigma$ excess in the $120$--$160\gev$ range of 
the dijet mass distribution in the $Wjj$ sample at the Tevatron, if real, signify  the existence of 
new physics at the electroweak scale.  We constructed a simple, weakly-coupled renormalizable theory 
with one multiplet of scalar particles obeying the 
$\left( \prod_{i=1}^3 \U(1)_{q_{\mathrm{L}i}} \times \U(1)_{u_{\mathrm{R}i}} \right)\! 
\times \U(3)_d \times \Z_3$ flavor 
symmetry, which ensures that the only source of flavor/CP violations is $V_\text{CKM}$.  We showed 
that the model can explain the two anomalies in terms of a single mass and a single coupling constant, 
without conflicting with existing bounds.
  
The flavor symmetry of the model also makes definite predictions on the amount of forward-backward 
asymmetries in $c\bar{c}$ and $b\bar{b}$.  The $c\bar{c}$ asymmetry arises predominantly from the 
$u\bar{u}$ initial state via the $t$-channel exchange of $\Phi_3^0$, while the $b\bar{b}$ asymmetry 
is dominated by $c\bar{c}$ via $\Phi_1^-$ exchange.  Therefore, the $c\bar{c}$ asymmetry is
predicted to be comparable to the $t\bar{t}$ asymmetry, while the $b\bar{b}$ asymmetry is expected 
to be suppressed due to the smallness of the $c$ parton distribution function. 

As mentioned in Sec.~\ref{sec:bump}, we also have a $Zjj$ production via 
$u\bar{s} \to Z \Phi_3^+ \to Zjj$, with smaller cross-section than $Wjj$.  For the LHC with 
$\sqrt{s} = 7\tev$, the production cross-sections of $Wjj$ and $Zjj$ are $47\pb$ and 
$12\pb$ respectively for the $160\gev$ resonances, and are $20\pb$ and 
$8\pb$ respectively for the $220\gev$ resonances. 

Finally, note that we get a sizeable forward-backward asymmetry because of $\mathcal{O}(1)$ coupling 
of the top quark with scalars of masses comparable to top mass, and as a result, we expect to see 
larger $t$-$\bar{t}$ production cross section that in the SM at high values of the invariant mass 
of the $t$-$\bar{t}$ pairs.  This would certainly be an interesting feature to observe at the LHC.

\section*{Acknowledgments}
TSR acknowledges stimulating discussions with A.~Martin.  This project  also benefited from 
conversations with S.~Ellis,  M.~Schmaltz, and B.~Tweedie.  The work of AEN and TSR was supported, 
in part, by the US Department of Energy under contract numbers DE-FGO2-96ER40956.

{\bf[Note added]} While this manuscript was in progress, 
Refs.~\cite{Yu:2011cw, 
Eichten:2011sh, Kilic:2011sr, Wang:2011uq, Cheung:2011zt} discussing the CDF 
$Wjj$ excess (but not the $t\bar{t}$ asymmetry) appeared in arXiv.  

{\bf[Note added 2]}
A week after the first version of our preprint appeared
on arXiv, we have received Ref.~\cite{Zhu:2011ww}, which claims that
the $\Phi_-$ mass has to be heavier than $540\gev$ to avoid too large a rate for
$B^+ \to K^+ \pi^0$, and that this would make our $Wjj$ excess
signal go away.  These claims, however, are incorrect.  First, even
granting $540\gev$ for the $\Phi^-$ mass, it is not true that the signal
would disappear, since it comes from the production of the scalar
$\Phi^0$ {\em directly from SM quarks,} whereas
Ref.~\cite{Zhu:2011ww} misunderstands that the signal is from an
$s$-channel $\Phi^-$ decaying to a $\Phi^0$ and a $W^-$.

Second, the $B^+ \to K^+\pi^0$ (i.e.~$\bar{b} \to d\bar{d}\bar{s}$) 
calculation of Ref.~\cite{Zhu:2011ww} is actually
technically incorrect.  The essential error is the
misidentification of the first term of 4-fermion operators~(\ref{eq:suckers}) 
as the standard QCD penguin operator $O_6$, where the latter is 
summed over all (active) quark flavors while the former is {\em not}.
As a result, dangerous decays such as $\bar{b} \to d\bar{d}\bar{s}$ 
are only generated via renormalization group running and are small. 
A proper calculation analyzing $\overline{B}^0 \to \pi^+ K^-$ (i.e.~$b \to u\bar{u} s$) 
was recently performed by Ref.~\cite{Blum:2011fa}, which found that the rate for this process
is enhanced by two orders of magnitude in our model.  

We propose two ways to avoid this constraint while keeping our signals intact.
The first is simply to make $\Phi^-$ heavier by a factor of a few to suppress 
operators~(\ref{eq:suckers}).  This will enhance the $T$ parameter~(\ref{eq:Tparam}) but it 
can be tuned to be small by adding, for example, 
an electroweak-triplet scalar with a nonzero vacuum expectation value which contributes negatively 
to $T$. 

The second way, which would require no tuning, is to enlarge the $\Z_3$ symmetry of the flavor 
symmetry~(\ref{eq:symmetry}) to the full permutation group $S_3$.  This amounts to the following.
To keep track of $S_3$ more easily, we rename $\Phi_{1,2,3}$ as $\Phi_{32,13,21}$, respectively,
and introduce $\Phi_{23,31,12}$ with the same gauge quantum numbers as $\Phi_{32,13,21}$ 
and the $\left( \prod_{i=1}^3 \U(1)_{q_{\mathrm{L}i}} \times \U(1)_{u_{\mathrm{R}i}} \right)$
charges in the manner obvious from the notation (e.g., $\Phi_{23}$ has charges $+1$ and $-1$
under $\U(1)_{q_{\mathrm{L}2}}$ and $\U(1)_{u_{\mathrm{R}3}}$, respectively, which is the opposite
of $\Phi_{32}$).  The six $\Phi$ fields form a sextet of $S_3$, and thus have a common mass 
$m_\Phi$ and a $\Phi^0$-$\Phi^-$ mass splitting parameter $\xi$ in the 
lagrangian~(\ref{eq:lagrangian}).  The Yukawa couplings in Eq.~(\ref{eq:lagrangian}) has to be
generalized to be $S_3$ invariant, i.e.,%

\begin{equation}
\cL_\text{Yukawa} = \lambda
\sum_\text{all permutations} \overline{q}_{\mathrm{L}1} \Phi_{12} u_{\mathrm{R}2} + \cc \,.
\end{equation} 
Then, instead of the dangerous operator 
$(\bar{q}_{\mathrm{L}2} u_{\mathrm{R}}) (\bar{u}_{\mathrm{R}} q_{\mathrm{L}2})$
(the first one of Eq.~(\ref{eq:4fermi})), we generate%
\begin{equation}
\begin{split}
{}&
(\bar{q}_{\mathrm{L}2} u_{\mathrm{R}}) (\bar{u}_{\mathrm{R}} q_{\mathrm{L}2})
+(\bar{q}_{\mathrm{L}3} u_{\mathrm{R}}) (\bar{u}_{\mathrm{R}} q_{\mathrm{L}3})
\\
= {}& 
-(\bar{q}_{\mathrm{L}1} u_{\mathrm{R}}) (\bar{u}_{\mathrm{R}} q_{\mathrm{L}1})
+\sum_{i=1}^3 (\bar{q}_{\mathrm{L}i} u_{\mathrm{R}}) (\bar{u}_{\mathrm{R}} q_{\mathrm{L}i})
\,,
\end{split}
\end{equation} 
in the gauge basis.  Going to the mass basis, the 2nd term will remain flavor-diagonal by the 
unitarity of $V$, hence not contributing to $b \to u\bar{u} s$, while the 1st term gives 
the $b$-dependent operator
\begin{equation}
  V_{ub}^* V_{ui} (\bar{b}_{\mathrm{L}} u_{\mathrm{R}}) (\bar{u}_{\mathrm{R}} d_{\mathrm{L}i})  \,.
\end{equation}
Remarkably, this is much smaller than the dangerous operator in Eq.~(\ref{eq:suckers}), 
which is only suppressed by $V_{cb}^* V_{ci}$.
Therefore,   the $b \to u\bar{u} s$ transition
rate in this new model is   suppressed by the same CKM factors as the
tree level standard model contribution, and is sufficiently small even
with a fairly light $\Phi^-$.  The doubling of $\Phi$ will increase the dijet cross-section
but still within the uncertainties discussed in section~\ref{sec:constraints}, especially due to
the fact that $\la$ in this model is smaller for the same values of the $t\bar{t}$ asymmetry.
A detailed study of this model is under investigation.


\end{document}